# Associated Lamé and various other new classes of elliptic potentials from sl(2,ℝ) and related orthogonal polynomials

**ASISH GANGULY**
**Department of Applied Mathematics**
**University of Calcutta**
**92 Acharya Prafulla Chandra Road**
**Kolkata - 700009**
**India**

**Abstract :** Using representations of sl(2,ℝ) generators which yield associated Lamé Hamiltonian we obtain new classes of elliptic potentials. We explicitly calculate eigenstates and spectra for these potentials and construct the associated orthogonal polynomials. We show that in the proper limit these potentials reduce to well-known exactly solvable potentials.

**e-mail :** asish@cucc.ernet.in

# 1. Introduction

The study of band-edge energies and wave functions for the class of periodic potentials plays an important role in quantum mechanics [1-6]. Recently, it was shown [1] within the framework of supersymmetric quantum mechanics [2] that the bosonic and fermionic sectors for a periodic superpotential $W(x)$ possess an identical spectra (including the zero mode) provided $\int_{\text{period}} W(x)\,dx = 0$. A wide class of solvable periodic potentials were obtained [3] by applying supersymmetry transformations on Lamé and associated Lamé equation [7,8] and it turned out that the superpartners of Lamé potentials are not physically identical except for some particular cases. On the other hand, periodic potentials belonging to the quasi-exactly solvable (QES) class have also been well studied in the literature [9-19]. It is useful to note that an early treatment on the non-compact $sl(2,\mathbb{R})$ Lie-algebraic scheme is due to Turbiner who formulated the basics of this formalism, gave the first classification of one-dimensional QES and proposed multidimensional generalizations. Indeed in Ref. [12] the connection between one-dimensional QES problems and the $sl(2,\mathbb{R})$ algebra was discovered (see also the review by Shifman [13]). Later in this formalism Lamé equation was algebraized [14,15] for integer as well as half-integer values of the potential parameter. Here it may be mentioned that a connection of periodic QES Hamiltonian within a compact Lie-algebraic approach was made by Alhassid et al [16] by expressing the Hamiltonian as a quadratic in su(2) generators.

Recently we have proposed [17] an algebraization of the associated Lamé equation namely

$$-\psi''(x) + [m(m+1)\,\mathrm{sn}^2x + \ell(\ell+1)\mathrm{cn}^2x/\mathrm{dn}^2x]k^2\,\psi(x) = E\psi(x) \qquad (1.1)$$

for m, $\ell \in \mathbb{R}$ and $(m,\ell)$ lying in at least one algebraic line (AL) in the $m-\ell$ plane. Here snx $\equiv$ sn(x,k), cnx $\equiv$ cn(x,k), dnx $\equiv$ dn(x,k) are three Jacobian elliptic functions of real modulus $k(0 < k^2 < 1)$ and $k'^2 = 1-k^2$ is the complementary modulus. An associated Lamé equation provides a more general class of periodic potentials due to the presence of two parameters $m,\ell$ and reduces to ordinary Lamé equation when either of $\ell$ and m takes the value 0 or $-1$. The explicit solutions were obtained by us for two cases : i) m, $\ell$ are both non-negative integers and ii) m is a half-integer and $\ell$ is an integer or half-integer. One of the aims of the present work is to extend our method of algebraization to the third new case when m is an integer but $\ell$ is a half-integer. To the best of our knowledge, our solutions of associated Lamé equation for integer and half-integer combinations of the parameters m and $\ell$ are new. We also generate two new classes of elliptic potentials and discuss their implications to the exactly solvable class. We also calculate band-edge eigenstates and energies for these new potentials.

The organization of the article is as follows. In section 2 we discuss the basic method of generating a wide range of families of elliptic potentials within $sl(2,\mathbb{R})$ thereby obtaining three new classes of potentials. In section 3 we algebraize the associated Lamé equation as a special case of one of them and make a systematic analysis to find the effective combination of two parameters m and $\ell$. Section 4 is devoted to the construction of associated orthogonal polynomials. In section 5 the solutions of associated Lamé equations are obtained for various values of m and $\ell$. Several new solutions are obtained here. In section 6 we calculate band-edge eigenstates and spectra for the two new classes of potentials obtained in section 2 and show their connection to the known exactly solvable potentials. Finally, our conclusions are presented in section 7.

## 2. Various families of elliptic potentials generated by sl(2,ℝ) algebra

QES models have been studied mainly from the point of view of two approaches. One approach is to find out a suitable Lie-algebraic representation of a QES Hamiltonian [12,13] and then to compute a finite part of the spectrum by diagonalising a matrix having finite block structure. A slight variation of this approach has been proposed in Ref. [19] where a class of QES rational potentials with normalizable zero energy state is generated from standard so(2,1) representation of exactly solvable model through a suitable coordinate transformation. Another approach, first suggested by Bender-Dunne [20,21], is to generate an orthogonal family of polynomials $\{P_j(E)\}$ in the energy variable E and then to find the roots of a critical polynomial $P_n(E)$ (where n is some positive integer) giving the energy eigenvalues of QES Hamiltonian. In fact the existence of such a critical polynomial is an evidence for a Hamiltonian to be QES.

Let us briefly recall the basic set-up within an sl(2,ℝ) Lie-algebra to generate the Schrödinger potential. The sl(2,ℝ) algebra is governed by the commutation relations

$$[T^+, T^-] = -2T^0, \qquad [T^0, T^\pm] = \pm T^\pm , \qquad (2.1)$$

where the three generators $T^\pm$, $T^0$ may be realized as

$$T^+ = \xi^2\partial_\xi - n\xi, \qquad T^0 = \xi\partial_\xi - \frac{1}{2} n, \qquad T^- = \partial_\xi , \qquad (2.2)$$

n being a non-negative integer. These generators act on polynomials in real variable $\xi$ of degree $\leq$ n.

The gauged Hamiltonian is taken as general quadratic combination of the generators in the form

$$H_G = -\sum_{a,b=0,\pm} C_{ab}\, T^a T^b - \sum_{a=0,\pm} C_a T^a - d , \qquad (2.3)$$

which, using (2.2) can be expressed as

$$H_G(\xi) = -\sum_{j=2}^{4} B_j(\xi)\,\partial_\xi^{j-2} \qquad (2.4)$$

$B_j(\xi)$ are the j-th degree polynomial in $\xi$ given by

$$B_4(\xi) = C_{++}\,\xi^4 + 2C_{+0}\,\xi^3 + C_{00}\xi^2 + 2C_{0-}\,\xi + C_{--}\,,$$

$$B_3(\xi) = 2(1-n)\,C_{++}\,\xi^3 + \{3(1-n)\,C_{+0} + C_+\}\,\xi^2 + \{(1-n)\,C_{00} + C_0\}\xi + (1-n)\,C_{0-} + C_-\,,$$

$$B_2(\xi) = n(n-1)\,C_{++}\,\xi^2 + n\{(n-1)\,C_{+0} - C_+\}\,\xi + \frac{n^2}{4}\,C_{00} - \frac{1}{2}\,nC_0 + d\,. \qquad (2.5)$$

Here the numerical parameters $C_{ij}$ (i,j=0,±) are symmetric with $C_{+-} = C_{-+} = 0$ and d is a suitably chosen constant.

Let us now convert $H_G$ to the form

$$H_G(x) = -\partial_x^2 + \left.\frac{B'_4 - 2B_3}{2\sqrt{B_4}}\right|_{\xi=\xi(x)} \partial_x - \left. B_2\right|_{\xi=\xi(x)}\,, \qquad (2.6)$$

through a coordinate transformation

$$x(\xi) = \int^{\xi} \frac{d\tau}{\sqrt{B_4(\tau)}}\,, \qquad (2.7)$$

where the prime denotes derivative with respect to $\xi$.

Now the Schrödinger Hamiltonian with potential V(x)

$$H(x)\,\psi(x) \equiv [-\partial_x^2 + V(x)]\,\psi(x) = E\,\psi(x)\,, \qquad (2.8)$$

can be gauge-transformed to the form

$$H_G(x)\,\chi(x) \equiv [-\partial_x^2 + 2A\,\partial_x + \frac{dA}{dx} - A^2 + V]\,\chi(x) = E\chi(x)\,, \qquad (2.9)$$

through an imaginary phase transformation

$$\psi(x) = e^{-\int A(x)dx}\,\chi(x) \qquad (2.10)$$

Comparing equations (2.6) and (2.9), the potential V(x) is obtained as

$$V(x) = \mathcal{A}^2 - \frac{d\mathcal{A}}{dx} - B_2 \Big|_{\xi=\xi(x)} , \quad (2.11)$$

where the gauge function $\mathcal{A}(x)$ is

$$\mathcal{A}(x) = \frac{B'_4 - 2B_3}{4\sqrt{B_4}} \Big|_{\xi=\xi(x)} \quad (2.12)$$

The Schrödinger potential V(x) can be written in terms of $B_j(\xi)$ in the following form

$$V(x) = \left[\frac{(B_4'-2B_3)(3B_4' - 2B_3)}{16B_4} - \frac{1}{4}(B_4'' - 2B_3' + 4B_2)\right]\Big|_{\xi=\xi(x)} \quad (2.13)$$

It is clear that the polynomial $B_4(\xi)$ determines the functional relation between $\xi$ and x through the equation (2.7). Table 2.1 contains all the twelve forms of $B_4(\xi)$ which relate $\xi$ to x through the elliptic functions [22,23]. Since our goal is to generate elliptic potentials these forms will facilitate our purpose. Note that once $B_4$ is chosen, equation (2.13) gives various families of elliptic potentials containing four parameters $C_+$, $C_-$, $C_0$ and n. In particular, for the transformation 1 of Table 2.1 we get a new class of periodic potentials

$$V(x) = P\,\mathrm{sn}^2x + Q\,\mathrm{sn}x\,\mathrm{cn}x + R\frac{\mathrm{sn}x\,\mathrm{cn}x}{\mathrm{dn}^2x} + S\frac{\mathrm{cn}^2x}{\mathrm{dn}^2x} , \quad (2.14)$$

where

$$P = \frac{k^2}{4}n(n+2) - \frac{C_0}{2}(n+1) + \frac{1}{4k^2}[C_0^2 - (C_+ - C_-)^2] ,$$

$$Q = \frac{1}{2k^2}(C_+ - C_-)[k^2(n+1) - C_0] ,$$

$$R = \frac{1}{2k^2}(C_+ - k'^2C_-)[k^2(n+1) + C_0] , \quad (2.15)$$

$$S = \frac{k^2}{4}n(n+2) + \frac{C_0}{2}(n+1) + \frac{1}{4k^2}[C_0^2 - \frac{1}{k'^2}(C_+ - k'^2C_-)^2] .$$

Here the constant d is chosen as

$$d = \frac{1}{4k^2}\,[C_-^{\,2} - (C_0^{\,2} + 2C_+C_-) + (\frac{C_+}{k'})^2] - \frac{n(n+2)}{2} \tag{2.16}$$

Again, for the second and third choices of $B_4(\xi)$ from Table 2.1, we are led to

$$V(x) = \begin{cases} k'^2\,[\dfrac{2}{cn^2x} - \dfrac{(n+3)\,(n+2)}{dn^2x}] & (2.17) \\[2ex] \dfrac{2}{sn^2x} - \dfrac{k'^2\,(n+3)\,(n+2)}{dn^2x} & (2.18) \end{cases}$$

where we have taken

$$C_+ = C_- = 0, \quad C_0 = \begin{cases} -k'^2(n+4) \\ \\ -(n+4) \end{cases}, \; d = \begin{cases} \dfrac{1}{4}\,[3n^2 + 12n + 8 - k^2\,(n^2 + 8n + 8), & (2.19) \\[2ex] \dfrac{1}{4}\,(3n^2 + 12n + 8) - \dfrac{k^2}{2}\,n\,(n+2). & (2.20) \end{cases}$$

Other choices of $B_4(\xi)$ can be similarly used. Later we will show that the families of elliptic potentials (2.14), (2.17) and (2.18) coincide with certain known potentials in the proper limit.

We now discuss the boundary conditions in corresponding spectral problem. We see that the Schrödinger equation (2.8) for the potential (2.14) can be defined over the entire x-axis. It follows from the oscillation theorem [8] that for a periodic potential of period 2K (where $K = \int_o^{\pi/2} d\alpha / \sqrt{1 - k^2 \sin^2\alpha}$) there exists a sequence $\{E_j\}_{j\geq 0}$ such that $E_o < E_1 \leq E_2 < E_3 \leq E_4 \ldots$, which are called the characteristic values of the energy parameter E and the periodic solutions of period 2K (or 4K) exist for $E = E_j$. The intervals of stability of the solutions are $(E_{2j}, E_{2j+1})$. For the 2K-periodic potentials (2.17) and (2.18) which are singular at $x = \pm K$ and $x = 0, 2K$ respectively, the domains of corresponding equations may be taken as (–K, K) and (0, 2K). Our choice of parameters guarantees that the wave functions vanish at boundary points.

## 3. Algebraization of Associated Lamé equation

Associated Lamé potentital (1.1) is a special case of the family of elliptic potentials (2.14) where

$$P = k^2 m(m+1)\,, \qquad Q = R = 0\,, \qquad S = k^2 \ell(\ell+1) \qquad (3.1)$$

Equation (3.1) gives the system of constraints determining the four parameters $C_+$, $C_-$, $C_0$ and n. Table 3.1 contains the exhaustive list of twelve sets of solutions for them. Since the spin parameter n is restricted to be a non-negative integer , the sixth column of Table 3.1 gives the corresponding restrictions on m, $\ell$ for validity of the solutions. Now twelve solutions for n represent four systems of parallel lines [see Fig.1] in m–$\ell$ plane: in the following these lines will be referred to as algebraic lines (ALs). Each point (m,$\ell$) on an i-th AL gives ($\eta_i$+1) algebraic eigenstates, where $\eta_i$ is a non-negative integer given by the numerical value of the spin parameter n associated with i-th AL. Clearly, if a point lies in the intersection of r ALs, we get $\sum_{i=1}^{r} (\eta_i+1)$ eigenstates, all of which are not necessarily distinct. (It should be mentioned that the oblique ALs are discontinuous at half-integer points that is, they do not meet axis-parallel ALs for the reason explained in section 4 [see Table 4.1]).

We now determine the effective region in m–$\ell$ plane. Denoting $p \equiv m(m+1)$ and $q \equiv \ell(\ell+1)$, we see that p, q are invariant under the translation m, $\ell \to m'$, $\ell'$ where $m' = -m-1$, $\ell' = -\ell-1$. Thus it is sufficient to take $\ell, m \geq -1/2$. In Fig.1 these two inequations give a square infinite region. Further due to the relations sn(x+2K) = –snx, sn(x+K) = cnx/dnx ($K = \int_0^{\pi/2} d\alpha / \sqrt{1 - k^2 \sin^2\alpha}$ is called complete elliptic integral of first kind), we can consider only $p \geq q$ without loss of generality. The restriction $p \geq q$ in turn implies that $m \geq \ell$. In Fig.1 this inequation gives a triangular infinite region in m–$\ell$ plane.

For any point outside this triangular region there exists a point in the region giving same associated Lamé potential. We call this region (shaded area in Fig. 1) as an effective region. Clearly the ALs lying entirely outside the effective region are not required for our purpose. It is easy to verify that those ALs correspond the solutions 7-12 of Table 3.1. It turns out that we need to consider first six solutions of Table 3.1 : indeed the solutions 7-12 can be generated from the solutions 1-6 by the translation $\ell, m \to \ell', m'$ where $\ell' = -\ell - 1$, $m' = -m-1$.

The analysis discussed so far reveals that the three-parameter family of associated Lamé potentials $V(x; m,\ell,k) = m(m+1)k^2\, sn^2x + \ell(\ell+1)\, k^2 cd^2x$ are generated† from $sl(2,\mathbb{R})$ algebra for $0 < k^2 < 1$ and $(m,\ell)$ lying in at least one AL in the effective region of $m-\ell$ plane. We call the points in the effective region not lying in any AL as critical points. For instance, though the point $(-1/2,-1/2)$ lies in the effective region, we cannot algebraize the corresponding associated Lamé equation $-\psi''(x) - (sn^2x + cd^2x)\, k^2\psi(x) / 4 = E\psi(x)$ in this scheme, as no ALs pass through this point. In the following sections we consider four cases namely when i) m and $\ell$ are both integers, ii) m is an integer and $\ell$ is a half-integer, iii) m and $\ell$ are both half integers and iv) m is a half-integer and $\ell$ is an integer.

The gauge Hamiltonian can be written in terms of $sl(2,\mathbb{R})$ generators in the form

$$H_G = H_{quadratic} + H_{linear}\,, \tag{3.2}$$

where

$$H_{quadratic} = -k'^2\, T^{+2} - (1+k'^2)\, T^{0\,2} - T^{-\,2}, \tag{3.3}$$

and for each of six solutions 1.-6. of Table 3.1, $H_{linear}$ is given in Table 3.2. It is to be noted that the generators $T^{\pm}$, $T^0$ are computed from (2.2) for the corresponding n in Table 3.2.

† we use the customary notation $cdx \equiv cnx/dnx$.

Before concluding the section, we wish to remark that the associated Lamé potential has two interesting limits :

$$V(x) \xrightarrow[k \to 0]{} 0 \qquad \text{( free particle)}$$

$$V(x) \xrightarrow[k \to 1]{} -\, m(m+1)\, \mathrm{sech}^2 x + \ell(\ell+1) + m(m+1) \qquad \text{( Pöschl-Teller potential)}$$

In above we have used the following limiting results :

$$\mathrm{sn}(x,k) \longrightarrow \begin{cases} \tanh x & k \to 1 \\ \sin x & k \to o \end{cases}, \qquad \mathrm{cn}(x,k) \longrightarrow \begin{cases} \mathrm{sech} x & k \to 1 \\ \cos x & k \to o \end{cases}, \qquad \mathrm{dn}(x,k) \longrightarrow \begin{cases} \mathrm{sech} x & k \to 1 \\ 1 & k \to o \end{cases} \qquad (3.4)$$

It may be mentioned that the present results are related to the ones obtained recently by Shifman and Turbiner [24]. Indeed the Hamiltonians 1.-2. of Table 3.2 possess the so-called energy-reflection symmetry.

# 4. Orthogonal polynomials in energy variable generated by eigenstates of a QES potential

It is well-known [15,20,21] that eigenfunctions of almost every one-dimensional QES Hamiltonian generate a family of orthogonal polynomials $\{P_j(E)\}$ in the energy variable E satisfying a three-term recursion relation of the form

$$P_{j+1}(E) = (\alpha_j E + \beta_j)\, P_j + \gamma_j\, P_{j-1}\,, \qquad j \geq 0 \qquad , \qquad (4.1)$$

where $\alpha_j$, $\beta_j$, $\gamma_j$ are independent of E, $\alpha_j \neq 0$, $\gamma_0=0$ and $\gamma_n=0$ for some positive integer n, provided we expand the algebraic eigenfunction (2.10) as power series in terms of suitable variable $y=y(\xi)$. Simple observation of equation (2.5) shows that we can separate the numerical parameters $C_{ij}$ and $C_i$ in $B_3(\xi)$ and $B_2(\xi)$ by rewriting them as

$$B_3(\xi) = \frac{1-n}{2}\, B_4'(\xi) + A_2(\xi)\,, \quad B_2(\xi) = \frac{n(n-1)}{12}\, B_4''(\xi) - \frac{n}{2} A_2'(\xi) + \frac{n(n+2)}{12}\, C_{00} + d\,, \qquad (4.2)$$

where $A_2(\xi)$ is a 2nd degree polynomial in $\xi$ given by

$$A_2(\xi) = C_+\xi^2 + C_0\xi + C_- \qquad (4.3)$$

Now due to the GL(2) symmetry, the form of $H_G(\xi)$ in (2.4) remains invariant under the coordinate transformation

$$\xi = \frac{ey + f}{gy + h} , \qquad \begin{pmatrix} e & f \\ g & h \end{pmatrix} \in GL(2,\mathbb{C}) , \qquad (4.4)$$

accompanied by the gauge transformation

$$\hat{\chi}(y) = \hat{\mu}(y)\, \chi\left(\frac{ey + f}{gy + h}\right) , \qquad \hat{\mu}(y) = (gy + h)^n \qquad (4.5)$$

The transformed Hamiltonian $\hat{H}_G(y)$ can be written as

$$-\hat{H}_G(y) = \hat{B}_4(y)\, \partial_y^2 + \left[\frac{1-n}{2} \hat{B}_4'(y) + \hat{A}_2(y)\right]\partial_y + \frac{n(n-1)}{12} \hat{B}_4''(y)$$
$$- \frac{n}{2} \hat{A}_2'(y) + \frac{n(n+2)}{12} \hat{C}_{00} + d_1 , \qquad (4.6)$$

where the prime now denotes derivative w.r.t. y. The transformed polynomials $\hat{B}_4(y)$ and $\hat{A}_2(y)$ are respectively of fourth and second degree and may be written as

$$\hat{B}_4(y) = \hat{C}_{++}\, y^4 + 2\, \hat{C}_{+0}\, y^3 + \hat{C}_{00}\, y^2 + 2\, \hat{C}_{0-}\, y + \hat{C}_{--} ,$$

$$\hat{A}_2(y) = \hat{C}_+\, y^2 + \hat{C}_0\, y + \hat{C}_- \qquad (4.7)$$

In (4.7) the transformed numerical parameters $\hat{C}_{ij}$, $\hat{C}_i$ are determined from the relations

$$\hat{B}_4(y) = \frac{(gy + h)^4}{(eh - fg)^2} B_4\left(\frac{ey + f}{gy + h}\right) , \qquad \hat{A}_2(y) = \frac{(gy + h)^2}{eh - fg} A_2\left(\frac{ey + f}{gy + h}\right) \qquad (4.8)$$

while the constant $d_1$ in (4.6) is given by

$$d_1 = d + \frac{n(n+2)}{12} (C_{00} - \hat{C}_{00}) \qquad (4.9)$$

Let us now consider the eigenvalue equation

$$\hat{H}_G(y)\,\hat{\chi}(y) = E\hat{\chi}(y)\ , \tag{4.10}$$

which is the transformed version of original gauged eigenvalue equation

$$H_G(\xi)\,\chi(\xi) = E\chi(\xi) \tag{4.11}$$

Here we have identified $\chi(\xi(x)) \equiv \chi(x)$ in (2.10). The transformed eigenfucntion $\hat{\chi}(y)$ may be taken as power series in y namely

$$\hat{\chi}(y) = \sum_{j=o}^{\infty} P_j(E)\,\frac{y^j}{j!} \tag{4.12}$$

It is straightforward to check that the coefficients $\{P_j(E)\}$, in general, satisfy a five-term recursion relation and to reduce this to an useful three-term recursion relation of the form (4.1), we require $\hat{C}_{++} = \hat{C}_{--} = 0$. This means that $\hat{B}_4(y)$ must vanish at 0 and $\infty$. It is clear from equation (4.8) that our problem remains to choose e, f, g, h in (4.4) in such a manner that the two distinct roots $\xi_1, \xi_2$ of $B_4(\xi)$ are mapped to 0, $\infty$ of $\hat{B}_4(y)$. Our choice corresponds to the set $e = \xi_2$, $f = -\xi_1$, $g = 1$, $h = -1$ so that $eh - fg = \xi_1 - \xi_2 \neq 0$. We are then led to a three-term recursion relation satisfied by $\{P_j(E)\}$

$$-[(2j-n+1)\,\hat{C}_{0-} + \hat{C}_-]P_{j+1} = [E + d_1 + \hat{C}_0\,(j - \frac{n}{2}) + \hat{C}_{00}\,(j - \frac{n}{2})^2\,]\,P_j$$
$$+\,j\,(j-1-n)\,[(2j-n-1)\,\hat{C}_{+0} + \hat{C}_+]\,P_{j-1}\,, \qquad j \geq 0\,, \tag{4.13}$$

where $P_{-1} \equiv 0$, $P_o \equiv 1$. Note that the relation (4.13) is of the form (4.1) provided that

$$(2j - n + 1)\,\hat{C}_{0-} + \hat{C}_- \neq 0\,, \qquad \forall j \geq 0 \tag{4.14}$$

We now examine whether (4.14) trivially holds or some additional restrictions need to be imposed on the spin parameter n. Clearly three cases are relevent which we discuss below.

**A. Case 1. Associated Lamé potential (1.1)**

This case corresponds to the transformation 1 of Table 2.1 where $B_4(\xi)$ has four distinct complex roots $\pm$ i , $\pm$ i/k′. Once the pair of the roots $\xi_1$, $\xi_2$ is chosen, the transformed numerical parameters $\hat{C}_{ij}$, $\hat{C}_i$ can be computed from (4.7) and (4.8) corresponding to each of the six solutions 1.–6. of Table 3.1 as follows :

*Solutions 1 and 2:* For the choice $\xi_1 = -i$, $\xi_2 = i$

$$\hat{C}_{++} = 0\,,\ \hat{C}_{+0} = k^2/2\,,\ \hat{C}_{00} = 2\,(k^2-2),\ \hat{C}_{0-} = k^2/2,\ \hat{C}_{--} = 0$$

$$\hat{C}_+ = \begin{cases} k^2(m-\ell)/2, \\ k^2(\ell+m+1)/2, \end{cases},\quad \hat{C}_0 = 0\,,\qquad \hat{C}_- = \begin{cases} k^2(\ell-m)/2, \\ -\,k^2\,(\ell+m+1)/2. \end{cases}$$

Then equation (4.14) gives

$$\frac{1}{2}\,k^2\,(2m-2j-1) \neq 0, \qquad \forall\, j \geq 0$$

We thus impose an additional restrictions on m as

$$m \neq \frac{1}{2}\,,\,\frac{3}{2}\,,\,\ldots \tag{4.15}$$

*Solutions 3 and 4:* For the choice $\xi_1 = \pm\, i/k'$, $\xi_2 = \mp\, i/k'$

$$\hat{C}_{++} = 0\,,\ \hat{C}_{+0} = -\,k^2/2\,,\ \ \hat{C}_{00} = 2(k^2-2)\,,\ \hat{C}_{0-} = -\,k^2/2\,,\ \hat{C}_{--} = 0$$

$$\hat{C}_+ = k^2\,(m-2\ell-\frac{1}{2})/2\,,\ \ \hat{C}_0 = (2\ell+1)\,(k^2-2)\,,\ \ \hat{C}_- = -\,k^2(2\ell+m+\frac{3}{2})/2$$

Then equation (4.14) gives

$$k^2\,(2j+2\ell+3)/2 \neq 0\,, \qquad \forall\, j \geq 0$$

This is always true in our effective region in m–$\ell$ plane (see Fig. 1).

*Solutions 5 and 6:* For the choice $\xi_1 = \pm i$, $\xi_2 = \mp i$

$$\hat{C}_{++} = 0,\ \hat{C}_{+0} = k^2/2\ ,\ \hat{C}_{00} = 2(k^2-2),\ \hat{C}_{0-} = k^2/2\ ,\ \hat{C}_{--} = 0$$

$$\hat{C}_{+} = k^2(2m - \ell + \frac{1}{2})/2\ ,\ \hat{C}_0 = (2m+1)(k^2-2)\ ,\ \hat{C}_{-} = k^2(2m + \ell + \frac{3}{2})/2$$

Then equation (4.14) gives

$$-k^2\,(2j + 2m + 3)/2 \neq 0\ , \qquad \forall\ j \geq 0$$

This is also true in our effective region.

**B. Case 2. Potential (2.17)**

This case corresponds to the transformation 2 of Table 2.1 and the solution (2.19). Here $B_4(\xi)$ has 4 distinct real roots $\pm$ 1, $\pm$ 1/k. For the choice $\xi_1 = 1$, $\xi_2 = -1$,

$$\hat{C}_{++} = 0,\ \hat{C}_{+0} = -k'^2/2\ ,\ \hat{C}_{00} = 2(1 + k^2)\ ,\ \hat{C}_{0-} = -k'^2/2\ ,\ \hat{C}_{--} = 0$$

and $$\hat{C}_{+} = \frac{k'^2}{2}(n+4)\ ,\quad \hat{C}_0 = 0,\quad \hat{C}_{-} = -\frac{k'^2}{2}(n+4)$$

Then equation (4.14) gives

$$-\frac{k'^2}{2}\,(2j+5) \neq 0\ , \qquad \forall j \geq 0 \tag{4.16}$$

**C. Case 3. Potential (2.18)**

This case corresponds to the transformation 3 of Table 2.1 and solution (2.10). The polynomial $B_4(\xi)$ has two real and two complex roots $\pm$ 1, $\pm$ ik′/k. For the choice $\xi_1 = 1$, $\xi_2 = -1$,

$$\hat{C}_{++} = 0,\ \hat{C}_{+0} = -1/2,\ \hat{C}_{00} = 2\,(1 - 2k^2)\ ,\ \hat{C}_{0-} = -1/2\ ,\ \hat{C}_{--} = 0$$

and $$\hat{C}_{+} = \frac{1}{2}(n+4)\ ,\quad \hat{C}_0 = 0,\quad \hat{C}_{-} = -\frac{1}{2}(n+4)$$

Then equation (4.14) gives

$$-\frac{1}{2}(2j+5) \neq 0\,, \qquad \forall j \geq 0 \tag{4.17}$$

Hence we have proved that in every case the eigenfunction generates a family of orthogonal polynomials, provided an additional restriction (4.15) is imposed on potential parameters for case 1 and the roots $\xi_1$, $\xi_2$ of the polynomial $B_4(\xi)$ are suitably chosen. The family $\{P_j(E)\}$ can be expressed in terms of monic polynomials $\{\tilde{P}_j(E)\}$ satisfying a recurrence relation of the type

$$\tilde{P}_{j+1} = (E-\lambda_j)\,\tilde{P}_j - \rho_j\,\tilde{P}_{j-1}\,, \tag{4.18}$$

$$\tilde{P}_j = \omega_j\,P_j\,, \qquad j \geq 0 \tag{4.19}$$

The explicit expressions of $\rho_j$, $\lambda_j$, $\omega_j$ together with the choice of roots ($\xi_1$, $\xi_2$) and overall restrictions on potential parameters for all cases are contained in Table 4.1. Note that the last two rows correspond to the potential (2.17) and (2.18). Further restriction (4.15) is reflected in Fig.1 showing discontinuities of oblique ALs at half-integer points. We see that $\rho_0 = 0$ and $\rho_{n+1} = 0$, so that the critical polynomial is $\tilde{P}_{n+1}$ and the energy eigenvalues of a QES Hamiltonian are precisely the zeros of $\tilde{P}_{n+1}$, provided all the zeros are real and simple.

Let us now turn to the construction of wave functions. From (2.10) and (4.5), the Schrödinger wave function $\psi(x)$ reads

$$\psi(x) = \mu(x)\,\frac{1}{\hat{\mu}(y)}\,\hat{\chi}(y)\Big|_{y\,=\,y(\xi(x))}\,, \tag{4.20}$$

where

$$\mu(x) = e^{-\int A(x)dx} \tag{4.21}$$

The gauge function $A(x)$ can be computed from (2.12) and by using (4.2)

$$\mu(x) = B_4^{-n/4}(\xi)\ \exp\left[\int^{\xi} \frac{A_2(\tau)}{2B_4(\tau)} d\tau\right]\Bigg|_{\xi=\xi(x)} \qquad (4.22)$$

Finally, the (n+1) eigenfunctions can be computed from (4.4), (4.5), (4.12) and (4.20) at the roots $E=E_i$ (i=0,1,…,n) of $\tilde{P}_{n+1}$ giving the final form

$$\psi_{E_i}(x) = \mu(x)\,(\xi(x) - \xi_2)^n \sum_{j=o}^{n} \frac{P_j(E_i)}{j!} \left(\frac{\xi(x)-\xi_1}{\xi(x)-\xi_2}\right)^j, \quad (i=0,1,\dots,n) \qquad (4.23)$$

It should be mentioned that the infinite power series expansion in (4.12) terminates after (n + 1) terms since the coefficients $P_j(E_i)$ vanishes for j > n.

## 5. Algebraic eigenfunctions and energy spectra of associated Lamé potential

Let us recall that here $B_4(\xi(x)) = dn^2x/cn^4x$ where $\xi = snx/cnx$ [transformation 1 of Table 2.1]. We have already obtained the explicit expression [cf. Eq. (4.23)] of eigenfunctions. It remains to calculate the gauge factor $\mu(x)$ for each of the six solutions 1.-6. of Table 3.1. From (4.22) and (4.3), $\mu(x)$ is given by

$$\mu(x)=\begin{cases} cn^{m+\ell}x\ dn^{-\ell}x & ,\quad n = m + \ell \qquad (5.1)\\ cn^{m-\ell-1}x\ dn^{\ell+1}x & ,\quad n = m - \ell - 1 \qquad (5.2)\\ \left.\begin{array}{l} cn^{m-1/2}x\ dn^{-\ell}x\ (cnx+ik'snx)^{\ell+1/2} \quad (5.3a)\\ cn^{m-1/2}x\ dn^{-\ell}x\ (cnx-ik'snx)^{\ell+1/2} \quad (5.3b)\end{array}\right\} & ,\quad n = m-\frac{1}{2}\\ \left.\begin{array}{l} cn^{\ell-1/2}x\ dn^{-\ell}x\ (cnx + isnx)^{m+1/2} \quad (5.4a)\\ cn^{\ell-1/2}x\ dn^{-\ell}x(cnx - isnx)^{m+1/2} \quad (5.4b)\end{array}\right\} & ,\quad n = \ell - \frac{1}{2} \end{cases}$$

Besides the elliptic modulus parameter $k(0 < k^2 < 1)$, the associated Lamé potential depends on two other real parameters m, $\ell$. In section 3 we have shown that it is sufficient to consider the points (m,$\ell$) in effective region (see Fig.1). The associated Lamé equation for any point (m,$\ell$) lying in at least one AL in the effective region thus gives an algebraic QES potential. In the following we shall confine ourselves to the case where m, $\ell \in (\mathbb{N} - 1) \cup (\mathbb{N} - 1/2)$. This can be decomposed into following four categories. It is to be noted that all points in the m-axis give Lamé potential.

## A. Case 1: m and $\ell$ are both non-negative integers

For $m \neq \ell$, the point (m, $\ell$) lies in two ALs belonging to two different systems given by $m+\ell=n$ and $m-\ell-1=n$ and for $m=\ell$, the point lies in one AL belonging to the system $m+\ell = 2m = n$, for n = 0, 1, 2, … . Each AL gives $(\eta_i + 1)$ algebraic eigenstates, where $\eta_i$ is the numerical value of the spin parameter n associated with the particular AL. Hence in either case, associated Lamé Hamiltonian possesses (2m+1) band-edge eigenstates corresponding to (2m+1) band-edge energies. This emphasizes the fact that the associated Lamé potential is a periodic QES and when m, $\ell$ are both non-negative integers there are m bound bands followed by a continuum band. The eigenstates for $m \neq \ell$ generate two distinct families of orthogonal polynomials of the form (4.18) where $\rho_j$, $\lambda_j$, $\omega_j$ are given by the entries 1 and 2 of Table 4.1. The solutions for m = 1,2 where $\ell$ is restricted to take (m+1) values 0, 1, …, m, have been obtained by us elsewhere [17] and it has been shown that the so-called Lamé potential is contained in the scheme as a particular case for $\ell$=0.

**B. Case 2: m and $\ell$ are both half an odd positive integer**

The point (m, $\ell$) lies in the intersection of two ALs belonging to two different systems given by m − 1/2 = n and $\ell$ − 1/2 = n. Since these two systems correspond to two pairs of algebraizations [see the solutions 3.-4. and 5.-6. of Table 3.1], the equations (4.23), (5.3) and (5.4) imply that each of the two algebraizations for each system gives (m + 1/2) and ($\ell$ + 1/2) complex eigenstates respectively which are conjugate to one another. Further the expressions of $\rho_j$, $\lambda_j$, $\omega_j$ [see the entries 3.-4. and 5.-6. of Table 4.1] imply that the complex eigenstates arising from two algebraizations of each system generate the same families of orthogonal polynomials in the energy variable E. It turns out that there are (m + 1/2) and ($\ell$ + 1/2) characteristic values of E for each of which we get two linearly independent solutions given by real and imaginary parts of the complex eigenstates. But we see that the solutions obtained from the system $\ell$ − 1/2 = n are included in the solutions obtained from the system m − 1/2 = n (note that m $\geq$ $\ell$ in the effective region). Hence when m and $\ell$ are both half-integers, there are (m + 1/2) characteristics values of E which are doubly degenerate. It may be mentioned that the point (m, $\ell$) in this case does not lie in the ALs belonging to the systems m+$\ell$=n and m−$\ell$−1=n due to the discontinuities of oblique ALs at half-integer points [see Fig.1]. The solutions for (1/2 , 1/2) , (3/2 , 1/2) and (3/2 , 3/2) have already been reported [17]. However, for some half-integer combinations of m, $\ell$, for instance (3/2 , 1/2) , we notice that only highest energy is doubly degenerate.

**C. Case 3. m is half an odd positive integer and $\ell$ is a non-negative integer**

The point (m, $\ell$) lies in one A.L. belonging to the system m − 1/2 = n. Thus, as before, there are (m + 1/2) characteristics values of E which are doubly degenerate. The solutions for (1/2 , 0), (3/2 , 0) and (3/2 , 1) are explicitely obtained by us in Ref. 17, and, the first two give Lamé potentials.

**D. Case 4. m is a non-negative integer and $\ell$ is half an odd positive integer**

This case is new and corresponds to the system $\ell$ − 1/2 = n. Two algebraizations given by the entries 5 and 6 of Table 4.1 yield ($\ell$ + 1/2) characteristics values of E for each of which there are two linearly independent solutions. Here $\ell$ takes values 1/2 , 3/2 , … and for each of them m is allowed to take an infinite set of values $\ell$ + 1/2 , $\ell$ + 3/2 , … in the effective region.

From the above discussions it is clear that in the effective region when $m \in \mathbb{N} \cup \mathbb{N} - 1/2$, $\ell$ is restricted to take (2m+1) values 0, 1/2, 1, 3/2 , …, m − 1/2, m. In the following examples $\phi_r(x)$ and $e_r$ denote the ordered levels of eigenstates and energy spectra. Note that in the half-integer cases, the parenthesized superscript in the eigenstate indicates the degeneracy of the eigenvalues.

We now consider the following examples.

(a) m=1

$\ell$ is allowed to take 3 values 0, 1/2, 1. Solutions for $\ell$ = 0, 1 were reported in [17]. The new case corresponds m = 1, $\ell$ = 1/2 and gives the associated Lamé potential as

$$V(x) = 2k^2 sn^2 x + \frac{3}{4} k^2 cd^2 x \qquad (5.5)$$

The wavefunctions and energy are

$$\phi_0^{(1)}(x) = \frac{\sqrt{1 + \text{cn}x}}{\sqrt{\text{dn}x}} (2\text{cn}x - 1), \quad \phi_0^{(2)}(x) = \frac{\text{sgn}(\text{sn}x)\sqrt{1-\text{cn}x}}{\sqrt{\text{dn}x}} (2\text{cn}x+1), \; e_0 = \frac{1}{4}(k^2+9)$$

(b) m=2

$\ell$ is allowed to take 5 values 0, 1/2, 1, 3/2, 2. The integer values of $\ell$ have already been given in Ref. 17.

i) $\ell = 1/2$ : Associated Lamé potential $V(x) = 6k^2\text{sn}^2x + \frac{3}{4} k^2\text{cd}^2x$ (5.6)

$$\phi_0^{(1)}(x)=\frac{\sqrt{1 + \text{cn}x}}{\sqrt{\text{dn}x}}(4\text{cn}^2x-2\text{cn}x-1), \qquad \phi_0^{(2)}(x)= \frac{\text{sgn}(\text{sn}x)\sqrt{1-\text{cn}x}}{\sqrt{\text{dn}x}}(4\text{cn}^2x+2\text{cn}x-1),$$

$$e_0= (k^2+25)/4$$

ii) $\ell = 3/2$ : Associated Lamé potential $V(x) = 6k^2\text{sn}^2x + \frac{15}{4} k^2\text{cd}^2x$ (5.7)

$$e_{0,1} = \frac{1}{4} [49 - 5k^2 + 2f_{\mp}(k)] , \qquad f_{\pm}(k) = 5(k^2-2) \pm \sqrt{k^4 + 25k'^2}$$

$$\phi_{0,1}^{(1)}(x) = \frac{\sqrt{1 + \text{cn}x}}{\text{dn}^{3/2}x} [8f_{\pm}(k)\text{cn}^3x - 4f_{\pm}(k)\text{cn}^2x - 2 \{2f_{\pm}(k) + 7k^2\}\text{cn}x + \{f_{\pm}(k)+7k^2\}]$$

$$\phi_{0,1}^{(2)}(x) = \frac{\text{sgn}(\text{sn}x)\sqrt{1-\text{cn}x}}{\text{dn}^{3/2}x} [8f_{\pm}(k)\, \text{cn}^3x + 4f_{\pm}(k)\text{cn}^2x-2\{2f_{\pm}(k)+7k^2\}\text{cn}x - \{f_{\pm}(k)+7k^2\}]$$

(c) m=3

$\ell$ is allowed to take 7 values 0, 1/2, 1, 3/2, 2, 5/2, 3.

i) $\ell = 0$ : Lamé potential $V(x) = 12k^2\text{sn}^2x$ (5.8)

$\phi_{0,4}(x) = \text{dn}x[5k^2\text{sn}^2x - f_{\pm}(k)]$ $\qquad e_{o,4} = 2f_{\mp}(k) + k^2$

$\phi_{1,5}(x) = \text{cn}x[5k^2\text{sn}^2x - g_{\pm}(k)]$ $\qquad e_{1,5} = 2g_{\mp}(k) + 1$

$\phi_{2,6}(x) = snx[5k^2sn^2x - \eta_\pm(k)]$ $\qquad$ $e_{2,6} = \eta_\mp(k) + 3(1+k^2)$

$\phi_3(x) = snx\ cnx\ dnx$ $\qquad$ $e_3 = 4(1 + k^2)$

where $f_\pm(k)=2k^2 + 1 \pm \sqrt{4k^4+k'^2}$, $g_\pm(k) = k^2 + 2 \pm \sqrt{4- k^2k'^2}$, $\eta_\pm(k)=2(k^2+1)\pm \sqrt{4k^4-7k^2+4}$

ii) $\quad \ell = 1/2$ : Associated Lamé potential $V(x) = 12k^2sn^2x + \frac{3}{4}k^2cd^2x$ $\qquad$ (5.9)

$$\phi_0^{(1)}(x) = \frac{\sqrt{1 + cnx}}{\sqrt{dnx}}[8cn^3x - 4cn^2x - 4cnx + 1) , \qquad e_0 = \frac{1}{4}(k^2 + 49)$$

$$\phi_0^{(2)}(x) = \frac{sgn(snx)\sqrt{1-cnx}}{\sqrt{dnx}}[8cn^3x + 4cn^2x - 4cnx - 1)$$

iii) $\quad \ell = 1$ : Associated Lamé potential $V(x) = 12k^2sn^2x + 2k^2cd^2x$ $\qquad$ (5.10)

$\phi_0(x) = cnx\ dn^2x$ $\qquad$ $e_0 = 1 + 4k^2$

$\phi_1(x) = snx\ dn^2x$ $\qquad$ $e_1 = 1 + 9k^2$

$$\phi_{2,3}(x) = \frac{snx\ cnx}{dnx}[sn^2x - \frac{1}{5k^2}(k^2 + 3 \pm \sqrt{k^4 + 9k'^2}] \qquad e_{2,3} = 10+2k^2 \mp 2\sqrt{k^4 + 9k'^2}$$

Remaining three eigenstates can be expressed as

$$\phi_i(x) = \frac{1}{dnx}[sn^4x - \frac{1}{10k^2}(9k^2+16-e_i)\ sn^2x + \frac{1}{15k^4}\{e_i^2 - 2(5k^2+18)e_i + (9k^4+156k^2+320)\}]$$

where the eigenvalues $e_i$ (i=4,5,6) are the roots of the cubic

$$e^3 - (11k^2 + 20)\ e^2 + (19k^4 + 216k^2 + 64)e - (9k^6 + 388k^4 + 448\ k^2) = 0$$

iv) $\quad \ell = 3/2$ : Associated Lamé potential $V(x) = 12k^2sn^2x + \frac{15}{4}k^2cd^2x$ $\qquad$ (5.11)

$$e_{0,1} = \frac{1}{4}[81 - 9k^2 + 2h_\mp(k)] \quad , \qquad h_\pm(k) = 7(k^2-2) \pm 2\sqrt{k^4 + 49k'^2}$$

$$\phi_{0,1}^{(1)}(x) = \frac{\sqrt{1 + cnx}}{dn^{3/2}x}[16h_\pm(k)cn^4x-8h_\pm(k)cn^3x-12\{h_\pm(k)+3k^2\}cn^2x+2\{2h_\pm(k)+9k^2\}cnx + \{h_\pm(k) + 9k^2\}]$$

$$\phi_{0,1}^{(2)}(x) = \frac{\mathrm{sgn}(\mathrm{sn}x)\sqrt{1-\mathrm{cn}x}}{\mathrm{dn}^{3/2}x}\,[16h_{\pm}(k)\mathrm{cn}^4x+8h_{\pm}(k)\mathrm{cn}^3x-12\{h_{\pm}(k)+3k^2\}\mathrm{cn}^2x$$
$$-2\{2h_{\pm}(k)+9k^2\}\mathrm{cn}x + \{h_{\pm}(k) + 9k^2\}]$$

v) $\ell=2$ : Associated Lamé potential $V(x) = 12k^2\mathrm{sn}^2x + 6k^2\mathrm{cd}^2x$ (5.12)

$$\phi_0(x) = \mathrm{dn}^3x \quad , \qquad e_0 = 9k^2$$

The three eigenstates can be expressed as

$$\phi_i(x) = \frac{\mathrm{cn}x}{\mathrm{dn}^2x}\,[\mathrm{sn}^4x - \frac{1}{10k^2}(4k^2+25-e_i)\,\mathrm{sn}^2x + \frac{1}{120k^4}\{e_i^2-2(2k^2+17)e_i-\frac{75}{2}k^4+156k^2+225\}]$$

where the three energies $e_i$ are the roots of the cubic

$$2e^3 - 2(8k^2 + 35)\,e^2 - (43k^4 - 640k^2 - 518)e + (525k^6 - 1365k^4 - 3312k^2 - 450) = 0$$

Remaining three energies $e_r$ are the roots of the cubic

$$2e^3 - 2(11k^2+35)e^2 - (37k^4-844k^2-518)e + 732k^6 - 1851k^4 - 4662k^2 - 450 = 0$$

and, the three eigenstates are given by

$$\phi_r(x) = \frac{\mathrm{sn}x}{\mathrm{dn}^2x}\,[\mathrm{sn}^4x - \frac{1}{10k^2}(9k^2+25-e_r)\,\mathrm{sn}^2x + \frac{1}{120k^4}\{e_r^2-2(5k^2+17)e_r-\frac{57}{2}k^4+306k^2+225\}]$$

vi) $\ell = 5/2$ : Associated Lamé potential $V(x) = 12k^2\mathrm{sn}^2x + \frac{35}{4}k^2\mathrm{cd}^2x$ (5.13)

Three eigenvalues $e_i$ are the roots of the cubic

$$64e^3-16(35k^2+179)\,e^2 + 4(39k^4+7398k^2+7459)e + (7915k^6-66871k^4-229507k^2-53361)=0$$

and, two linearly independent solutions for each $e_i$ are

$$\phi_i^{(1)}(x) = \frac{\sqrt{1 + \mathrm{cn}x}}{\mathrm{dn}^{5/2}x}\,[32\beta_i\mathrm{cn}^5x-16\beta_i\mathrm{cn}^4x-16(\gamma_i+2\beta_i)\mathrm{cn}^3x+4(2\gamma_i+3\beta_i)\mathrm{cn}^2x$$
$$+ 2(4\gamma_i+3\beta_i+792k^4)\mathrm{cn}x - (2\gamma_i+\beta_i+792k^4)]$$

$$\phi_i^{(2)}(x) = \frac{\mathrm{sgn}(\mathrm{sn}x)\sqrt{1-\mathrm{cn}x}}{\mathrm{dn}^{5/2}x}\,[32\beta_i\mathrm{cn}^5x+16\beta_i\mathrm{cn}^4x-16(\gamma_i+2\beta_i)\mathrm{cn}^3x-4(2\gamma_i+3\beta_i)\mathrm{cn}^2x$$
$$+ 2(4\gamma_i+3\beta_i+792k^4)\mathrm{cn}x + (2\gamma_i+\beta_i+792k^4)]$$

where $\gamma_i(k) = 22k^2 (37k^2 + 9 - 4e_i)$

$\beta_i(k) = 16e_i^2 - 8(27k^2 + 29)e_i + 845k^4 + 1966k^2 + 441$

vii) $\ell$=3 : Associated Lamé potential $V(x) = 12k^2 sn^2x + 12k^2 cd^2x$ (5.14)

Three band-edge eigenstates are given by

$$\phi_i(x)= \frac{snxcnx}{dn^3x}[sn^4x - \frac{1}{10k^2} (4k^2+36-e_i)\, sn^2x + \frac{1}{120k^4}\{e_i^2-4(k^2+13)e_i + 24(11k^2 + 24)\}]$$

(i = 0, 1, 2)

where the three eigenvalues $e_i$ are the roots of the cubic

$$e^3 - 8(k^2+7)e^2 + 16(k^4+35k^2+49)e - 192(7k^4+31k^2+12) = 0$$

In particular, for $k^2 = 2/3$, the energies and eigenvalues can be obtained in exact form :

$$\phi_0(x) = \frac{snx\, cnx}{dn^3x} [sn^4x - 4sn^2x + \frac{9}{2}] \qquad e_0 = 12$$

$$\phi_1(x) = \frac{snx\, cnx}{dn^3x} [sn^4x - 3sn^2x + \frac{3}{2}] \qquad e_1 = 56/3$$

$$\phi_2(x) = \frac{snx\, cnx}{dn^3x} [sn^4x - \frac{6}{5} sn^2x + \frac{3}{10}] \qquad e_2 = 92/3$$

Remaining four energies $e_r$ are the roots of the biquadratic

$$e^4 - 4(5k^2+14)e^3 + 2(59k^4+616k^2+392)e^2-12(15k^6+698k^4+1280k^2+192)e$$
$$+ 9k^2(9k^6+1824k^4+8320k^2+3072) = 0$$

and, corresponding four eigenstates can be expressed as

$$\phi_r(x) = \frac{1}{dn^3x}[sn^6x - \frac{1}{10k^2} (9k^2+36-e_r)sn^4x+ \frac{1}{120k^4}\{e_r^2-2(5k^2+26)e_r + 9k^4+480k^2+576\}sn^2x$$
$$+ \frac{1}{720k^6} \{e_r^3 - (11k^2+56)e_r^2 + (19k^4+716k^2+784)e_r - 3(3k^6+604k^4+2560k^2+768)\}]$$

Higher values of m and $\ell$ can be similarly considered, in principle. But, in practice, as m,$\ell$ increases, calculations become very complicated due to the fact that the analytical solutions of higher degree critical polynomials are quite lengthy. However it may be mentioned that our algebraization scheme discovers several new families of associated Lamé potentials for integer and half-integer combinations of m,$\ell$, e.g., the potentials (5.5)-(5.7), (5.9), (5.11) and (5.13), whose solutions can be written analytically.

## 6. Elliptic generalizations of Gendenshtein and other exactly solvable periodic potentials

In section 2 we have obtained two new families of potentials (2.17) and (2.18) for the second and third choices [ transformations 2.-3. of Table 2.1 ] of $B_4(\xi)$. For the choice $B_4(\xi) = (1-\xi^2)(1-k^2\xi^2)$, we get

$$V(x) = k'^2 \left[\frac{2}{cn^2 x} - \frac{(n+3)(n+2)}{dn^2 x}\right] \quad , \tag{6.1}$$

V(x) depends on two parameters $k(0<k^2<1,\ k'^2=1-k^2)$ and $n(\in \mathbb{N} -1)$. Further we choose the domain (-K,K) where 2K is the period of the potential and given by $K = \int_0^{\pi/2} d\alpha/\sqrt{1-k^2\sin^2\alpha}$. We may note that as $k \rightarrow 1, 0$ the domain reduces to $(-\infty, \infty)$ and $(-\pi/2, \pi/2)$. It is interesting to observe following two limits:

$$V(x) \xrightarrow[k\to 1]{} 0 \quad , \quad x \in (-\infty, \infty) \qquad \text{(free particle)} \tag{6.2}$$

$$V(x) \xrightarrow[k\to 0]{} 2\tan^2 x - (n+1)(n+4), \quad x \in (-\pi/2, \pi/2) \tag{6.3}$$

which is an exactly solvable periodic potential [25].

Again, for the choice $B_4(\xi) = (1-\xi^2)(k'^2+k^2\xi^2)$ [transformation 3. of Table 2.1], we get,

$$V(x) = \frac{2}{sn^2 x} - \frac{k'^2(n+3)(n+2)}{dn^2 x} \ , \quad x \in (0, 2K) \tag{6.4}$$

where n($\mathbb{N}$ −1) and k($0<k^2<1$) are two parameters. The following two limits of (6.4) are noted :

$$V(x) \xrightarrow[k\to 1]{} 2\operatorname{cosech}^2 x + 2 \quad , \qquad x \in (0, \infty) \tag{6.5}$$

which is singular Gendenshtein potential [26] and

$$V(x) \xrightarrow[k\to 0]{} 2\cot^2 x - (n+1)(n+4), \quad x \in (0, \pi) \tag{6.6}$$

which also belongs to the exactly-solvable class. Note that in above we have used the limiting results (3.4).

The elliptic potentials (6.1) and (6.4) are thus new families of periodic potentials belonging to QES class and can be considered as generalized versions of exactly solvable $\tan^2 x$, $\cot^2 x$ and $\operatorname{cosech}^2 x$ potentials. These potentials are generated from a standard homogeneous quadratic combination of sl(2,$\mathbb{R}$ ) generators plus a linear operator. The gauged Hamiltonian can be written explicitly in terms of sl(2,$\mathbb{R}$ ) generators as

Potential (6.1) :
$$H_G = -k^2 T^{+2} + (1+k^2)\, T^{0\,2} - T^{-2} + k'^2(n+4)T^0 - \frac{1}{4}\,[3n^2 + 12n + 8 - k^2\,(n^2 + 8n + 8)], \tag{6.7}$$

Potential (6.4) :
$$H_G = k^2 T^{+2} + (1-2k^2)T^{0\,2} - k'^2\, T^{-2} + (n+4)\, T^0 - \frac{1}{4}\,[3n^2 + 12n + 8 - 2k^2 n\,(n + 2)], \tag{6.8}$$

where the generators $T^{\pm}$, $T^0$ are computed from (2.2) for $n\in\mathbb{N} -1$. The gauge factor $\mu(x)$ is obtained from (4.22), (2.19) and (2.20) for the corresponding choice of $B_4(\xi)$ [transformation 2.-3. of Table 2.1] as

Potential (6.1) : $\mu(x) = cn^2x / dn^{n+2}x$ (6.9)

Potential (6.4) : $\mu(x) = sn^2x / dn^{n+2}x$ (6.10)

The (n+1) band-edge eigenstates and spectra can now be easily computed by constructing orthogonal polynomials using the entries 7, 8 of Table 4.1.

We now consider the following examples.

(a) The class of potentials given by equation (6.1)

i) n=0 : $$V(x) = 2k'^2 \left(\frac{1}{cn^2x} - \frac{3}{dn^2x}\right) \tag{6.11}$$

$\phi_o(x) = \dfrac{cn^2x}{dn^2x}$ , $e_0 = -2k'^2$

ii) n=1 : $$V(x) = 2k'^2 \left(\frac{1}{cn^2x} - \frac{6}{dn^2x}\right) \tag{6.12}$$

$\phi_o(x) = \dfrac{cn^2x}{dn^3x}$ , $e_0 = 7k^2 - 8$

$\phi_1(x) = \dfrac{snxcn^2x}{dn^3x}$ , $e_1 = 2k^2 - 3$

(b) The class of potentials given by equation (6.4)

i) n=0 : $$V(x) = \frac{2}{sn^2x} - \frac{6k'^2}{dn^2x} \tag{6.13}$$

$\phi_0(x) = \dfrac{sn^2x}{dn^2x}$ , $e_0 = -2$

ii) n=1 : $$V(x) = \frac{2}{sn^2x} - \frac{12k'^2}{dn^2x} \tag{6.14}$$

$\phi_o(x) = \dfrac{sn^2x}{dn^3x}$ , $e_0 = k^2 - 8$

$$\phi_1(x) = \frac{sn^2xcnx}{dn^3x} \qquad , \qquad e_1 = k^2 - 3$$

Higher values of m can be similarly considered. We have thus found two new families of QES periodic potentials which in the proper limit reduce to well-known exactly-solvable class.

## 7. Conclusion

In this paper we have systematically studied the basic method of generating various classes of elliptic potentials within an sl(2,$\mathbb{R}$) Lie-algebraic formulation. We have given a wide range of choice of coordinate transformation in Table 2.1. Using the transformation 1.-3. of Table 2.1 we have obtained three new classes of QES periodic potentials, which, in the limits $k \to 0,1$, reduce to Pöschl-Teller, singular Gendenshtein and other known exactly solvable class. Other choices from Table 2.1 can be similarly used to generate various new classes of elliptic potentials. Further, we have shown that the associated Lamé potential can be derived as a special case. This certainly adds an important member in the list of algebraic QES. We have explicitly obtained the eigenstates and spectra for these potentials. In fact our algebraic approach discovers a new class of associated Lamé potential when either of the two parameters m and $\ell$ are integers and other half-integers. It is shown that the eigenfunctions of our QES Hamiltonians generate a family of orthogonal polynomials {Pj(E)} satisfying a three term recurrence relation of the form (4.18). It is to be mentioned that the coefficients $\rho_j$ are not strictly positive for $1 \leq j \leq n$, in contrast to the cases discussed in Ref. 15. However, the (n+1) roots of the critical polynomial are all real and simple.

## Acknowledgement

I would like to thank Professor B Bagchi for guidance. I also like to thank the referee for constructive criticisms.

## References

[1]G. Dunne and J. Feinberg, Phys. Rev. **D57**, 1271 (1998).

[2] B. Bagchi, *Supersymmetry in quantum and classical mechanics* (Chapman & Hall / CRC, Boca Raton, 2000).

[3]A. Khare and U. Sukhatme, J. Math. Phys. **40**, 5473 (1999).

[4]A.R. Its and V.B. Matveev, Theor. Matem. Fiz. **23**, 51 (1975) (in Russian).

[5]A. Treibich and J.L. Verdier, Compt. Rendus Acad. Sci. Paris. **311**, 51 (1990).

[6] E.D. Belokolos, A.I. Bobenko, V.Z. Enol′skii and A. R. Its, *Algebro-Geometreic Approach to Nonlinear Integrable Equations* (Springer-Verlag, Berlin, 1994), Chap. 7.7.

[7] F.M. Arscott, *Periodic Differential Equations* (Pergamon, Oxford, 1964).

[8] W. Magnus and S. Winkler, *Hill's Equation* (Wiley, New York, 1966).

[9]N. Kamran and P.J. Olver, J. Math. Anal. Appl. **145**, 342 (1990).

[10]A. González-López, N. Kamran and P.J. Olver, Contemp. Math. **160**, 113 (1994).

[11]M. Razavy, Am. J. Phys. **48**, 285 (1980).

[12]A.V. Turbiner, Commun. Math. Phys. **118**, 467 (1988).

[13]M.A. Shifman, Contemp. Math. **160**, 237 (1994).

[14]A. V. Turbiner, J. Phys. **A22**, L1 (1989).

[15]F. Finkel, A. González-López and M.A. Rodríguez, J. Math. Phys. **37**, 3954 (1996); J.Phys. **A33**, 1519 (2000).

[16]Y. Alhassid, F. Gürsey and F. Iachello, Phys. Rev. Lett. **50**, 873 (1983).

[17]A. Ganguly, Mod. Phys. Lett **A15**, 1923 (2000). It is to be noted that the expression for $\phi_1^{(1)}(x)$ for m = 3/2, $\ell$ = 1/2 has to be read as $\phi_1^{(1)}(x) = (2\mathrm{sn}^2 x - 1) / \sqrt{\mathrm{dn} x}$.

[18]M.A. Shifman, Int. J. Mod. Phys. **A4**, 2897 (1989).

[19]B. Bagchi and C. Quesne, Phys. Lett. **A230**, 1 (1997).

[20]C.M. Bender and G.V. Dunne, J. Math. Phys. **37**, 6 (1996).

[21]A. Krajewska, A. Ushveridze and Z. Walczak, Mod. Phys. Lett. **A12**, 1131 (1997).

[22] E.T. Whittaker and G.N. Watson, *A Course of Modern Analysis* (Cambridge University Press, Cambridge, 1989), Chap. 18-23.

[23] I.S. Gradshteyn and I.M. Ryzhik, *Table of Integrals, Series and Products* (Academic, New York, 1980). Chap. 6 and 8.

[24]M.A. Shifman and A.V. Turbiner, Phys. Rev. **A59**, 1791 (1999).

[25]G. Lévai, J. Phys. **A22**, 689 (1989).

[26] L.E. Gendesnhtein Zh. Eksp. Theor. Fiz. Pis. Red. **38**, 299 (1983) [JEPT Lett. **38**, 356 (1983)].

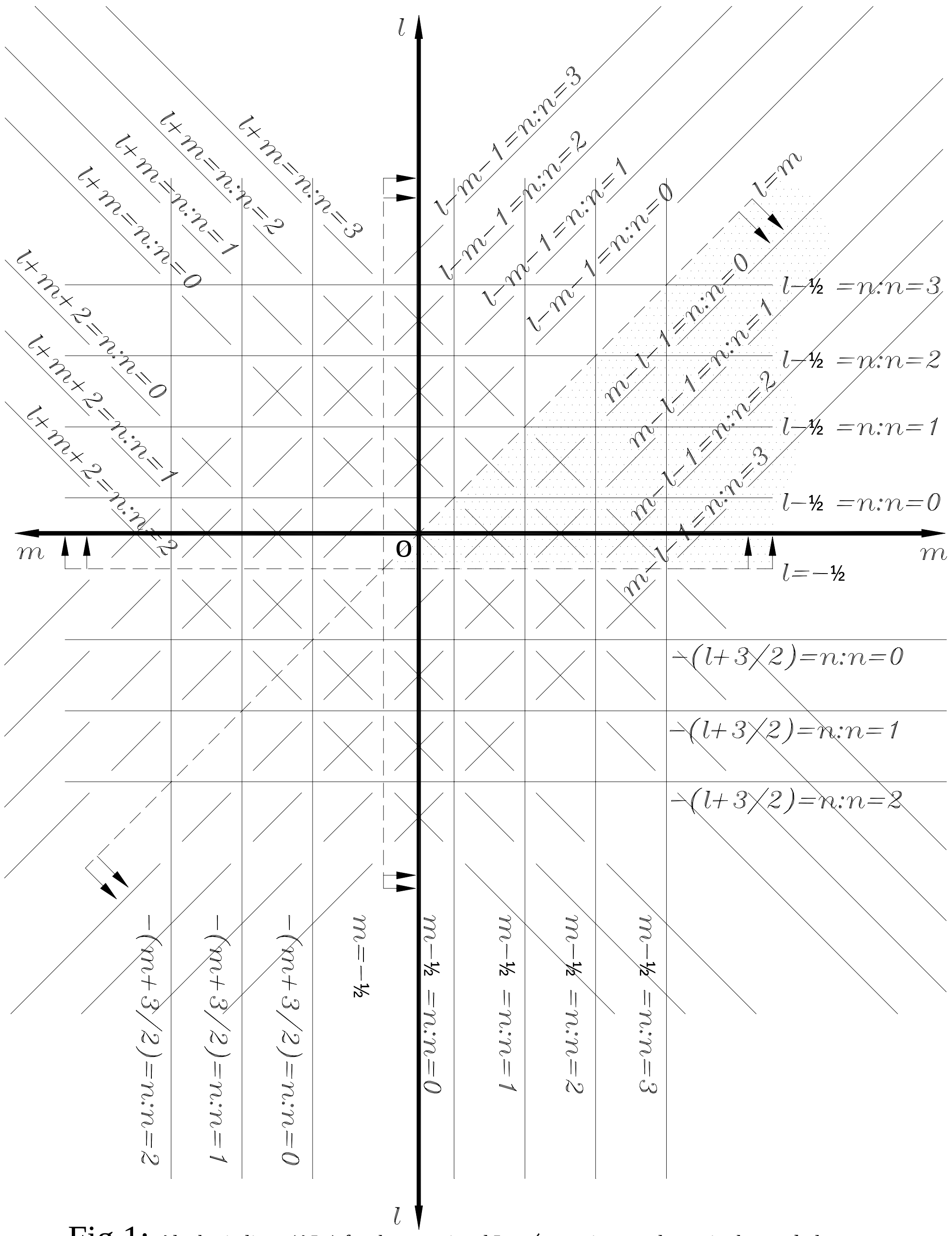

Fig.1: Algebraic lines (ALs) for the associated Lamé equation are drawn in the $m$-$l$ plane for some non-negative integer values of n. Three dashed lines determine a triangular infinite region as effective region (shaded area). Oblique ALs are discontinuous at half- integer points.

**Table 2.1**

| Sl. No. | $B_4(\xi)$ | Zeros of $B_4(\xi)$ | $\xi_o$ | $x(\xi) = \int_{\xi_o}^{\xi} B_4^{-1/2}(\tau)\, d\tau$ gives $\xi = \xi(x)$ | $B_4(\xi(x))$ |
|---|---|---|---|---|---|
| 1. | $(1 + \xi^2)(1 + k'^2\xi^2)$ | $\pm i, \pm i/k'$ | 0 | snx/cnx | $dn^2x/cn^4x$ |
| 2. | $(1 - \xi^2)(1 - k^2\xi^2)$ | $\pm 1, \pm 1/k$ | 0 | snx | $cn^2x\, dn^2x$ |
| 3. | $(1 - \xi^2)(k'^2 + k^2\xi^2)$ | $\pm 1, \pm ik'/k$ | 1 | –cnx | $sn^2x\, dn^2x$ |
| 4. | $(1 - \xi^2)(\xi^2 - k'^2)$ | $\pm 1, \pm k'$ | 1 | –dnx | $k^4sn^2x\, cn^2x$ |
| 5. | $(1 + k^2\xi^2)(1 - k'^2\xi^2)$ | $\pm i/k, \pm 1/k'$ | 0 | snx/dnx | $cn^2x/dn^4x$ |
| 6. | $(\xi^2-1)(\xi^2-k^2)$ | $\pm 1, \pm k$ | $-\infty$ | –1/snx | $cn^2xdn^2x/sn^4x$ |
| 7. | $(\xi^2-1)(k'^2\xi^2 + k^2)$ | $\pm 1, \pm ik/k'$ | 1 | 1/cnx | $sn^2xdn^2x/cn^4x$ |
| 8. | $(\xi^2-1)(1-k'^2\xi^2)$ | $\pm 1, \pm 1/k'$ | 1 | 1/dnx | $k^4sn^2xcn^2x/dn^4x$ |
| 9. | $(1+\xi^2)(k'^2+\xi^2)$ | $\pm i, \pm ik'$ | $-\infty$ | –cnx/snx | $dn^2x/sn^4x$ |
| 10. | $(\xi^2+k^2)(\xi^2-k'^2)$ | $\pm ik, \pm k'$ | $-\infty$ | –dnx/snx | $cn^2x/sn^4x$ |
| 11. | $(1-\xi^2)(1-k^2\xi^2)$ | $\pm 1, \pm 1/k$ | –1 | –cnx/dnx | $k'^4sn^2x/dn^4x$ |
| 12. | $(\xi^2-1)(\xi^2-k^2)$ | $\pm 1, \pm k$ | –1 | dnx/cnx | $k'^4sn^2x/cn^4x$ |

Different coordinate transformations based on the choice of $B_4(\xi)$ and path of integration giving $\xi=\xi(x)$ in terms of three Jacobi elliptic functions snx≡sn(x,k), cnx≡cn(x,k) and dnx≡dn(x,k) of real modulus $k(0<k^2<1, k'^2=1-k^2)$ are given.

**Table 3.1**

| **Sl. No.** | **n** | **$C_+$** | **$C_-$** | **$C_o$** | **Restrictions on m, $\ell$** |
|---|---|---|---|---|---|
| 1. | $\ell+m$ | 0 | 0 | $k^2(\ell-m)$ | $\ell+m \in \mathbb{N} -1$ |
| 2. | $m-\ell-1$ | 0 | 0 | $-k^2(\ell+m+1)$ | $m-\ell \in \mathbb{N}$ |
| 3. | $m - 1/2$ | $ik'(2\ell+1)$ | $ik'(2\ell+1)$ | $-k^2(m+1/2)$ | $m \in \mathbb{N} - 1/2, \ell \in \mathbb{R}$ |
| 4. | $m - 1/2$ | $-ik'(2\ell+1)$ | $-ik'(2\ell+1)$ | $-k^2(m+1/2)$ | $m \in \mathbb{N} - 1/2, \ell \in \mathbb{R}$ |
| 5. | $\ell - 1/2$ | $ik'^2(2m+1)$ | $i(2m+1)$ | $k^2(\ell+1/2)$ | $\ell \in \mathbb{N} - 1/2, m \in \mathbb{R}$ |
| 6. | $\ell - 1/2$ | $-ik'^2(2m+1)$ | $-i(2m+1)$ | $k^2(\ell+1/2)$ | $\ell \in \mathbb{N} - 1/2, m \in \mathbb{R}$ |
| 7. | $-(\ell + m + 2)$ | 0 | 0 | $-k^2(\ell-m)$ | $-(\ell+m) \in \mathbb{N} + 1$ |
| 8. | $\ell - m - 1$ | 0 | 0 | $k^2(\ell+m+1)$ | $\ell - m \in \mathbb{N}$ |
| 9. | $-(m + 3/2)$ | $ik'(2\ell+1)$ | $ik'(2\ell+1)$ | $k^2(m+1/2)$ | $- m \in \mathbb{N} + 1/2, \ell \in \mathbb{R}$ |
| 10. | $-(m + 3/2)$ | $-ik'(2\ell+1)$ | $-ik'(2\ell+1)$ | $k^2(m+1/2)$ | $- m \in \mathbb{N} + 1/2, \ell \in \mathbb{R}$ |
| 11. | $-(\ell+3/2)$ | $ik'^2(2m+1)$ | $i(2m+1)$ | $-k^2(\ell+1/2)$ | $- \ell \in \mathbb{N} + 1/2, m \in \mathbb{R}$ |
| 12. | $-(\ell+3/2)$ | $-ik'^2(2m+1)$ | $-i(2m+1)$ | $-k^2(\ell+1/2)$ | $- \ell \in \mathbb{N} + 1/2, m \in \mathbb{R}$ |

The solutions of a system of four non-linear equations for four parameters n, $C_\pm$ and $C_o$ given by (3.1) are provided. Proper restrictions on $m,\ell$ are essential to restrict the spin parameter n to a non-negative integer. It is interesting to note that the solutions 7–12 can be obtained from 1–6 under the translations $m \to m' = -m-1$, $\ell \to \ell' = -\ell-1$.

**Table 3.2**

| Sl No. | n | $H_{linear}$ |
|---|---|---|
| 1. | $m + \ell$ | $-k^2(\ell - m)\, T^0 + \frac{k^2}{4}(\ell - m)^2 + \frac{n(n+2)}{2}$ |
| 2. | $m - \ell - 1$ | $k^2(\ell+m+1)\, T^0 + \frac{k^2}{4}(\ell+m+1)^2 + \frac{n(n+2)}{2}$ |
| 3. | $m - 1/2$ | $-ik'(2\ell+1)\,(T^+ + T^-) + k^2\,(m + \frac{1}{2})\, T^0 + \frac{(2\ell+1)^2}{4} + \frac{k^2}{4}(m + \frac{1}{2})^2 + \frac{n(n+2)}{2}$ |
| 4. | $m - 1/2$ | $ik'(2\ell+1)\,(T^+ + T^-) + k^2(m + \frac{1}{2})\, T^0 + \frac{(2\ell+1)^2}{4} + \frac{k^2}{4}(m + \frac{1}{2})^2 + \frac{n(n+2)}{2}$ |
| 5. | $\ell - 1/2$ | $-i(2m+1)\,(k'^2\, T^+ + T^-) - k^2\,(\ell + \frac{1}{2})\, T^0 + \frac{(2m+1)^2}{4} + \frac{k^2}{4}(\ell + \frac{1}{2})^2 + \frac{n(n+2)}{2}$ |
| 6. | $\ell - 1/2$ | $i(2m+1)\,(k'^2\, T^+ + T^-) - k^2(\ell + \frac{1}{2})\, T^0 + \frac{(2m+1)^2}{4} + \frac{k^2}{4}(\ell + \frac{1}{2})^2 + \frac{n(n+2)}{2}$ |

Lie-algebraic representations of the linear part $H_{linear}$ of assocaited Lamé Hamiltonian $H_G$ are given for different solutions of n, in terms of the three sl(2,$\mathbb{R}$ ) generators $T^{\pm}$, $T^0$. The quadratic part is given by $H_{quadratic} = -k'^2\, T^{+2} - (1 + k'^2)\, T^{02} - T^{-2}$.

## Table 4.1

| Sl No | n | $\rho_j$ | $\lambda_j$ | $\omega_j$ | $(\xi_1,\xi_2)$ | Restrictions on $m,\ell$ |
|---|---|---|---|---|---|---|
| 1. | $m+\ell$ | $(\frac{1}{2}k^2)^2 j(\ell+m+1-j)(2j-2\ell-1)(2m-2j+1)$ | $\frac{1}{2}k^2\{\ell(\ell+1)+m(m+1)\}+\frac{1}{2}(2-k^2)(\ell+m-2j)^2$ | $\dfrac{(\frac{k^2}{2})^j \prod_{i=0}^{m-1}(2m-2i-1)}{\prod_{i=0}^{m-1-j}(2m-2j-2i-1)}$ | $(-i, i)$ | $m\notin\mathbb{N}-\frac{1}{2}, \ell\in\mathbb{R}$<br>$m+\ell\in\mathbb{N}-1$ |
| 2. | $m-\ell-1$ | $(\frac{1}{2}k^2)^2 j(m-\ell-j)(2m-2j+1)(2j+2\ell+1)$ | $\frac{1}{2}k^2\{\ell(\ell+1)+m(m+1)\}+\frac{1}{2}(2-k^2)(m-\ell-1-2j)^2$ | ″ | $(-i,i)$ | $m\notin\mathbb{N}-\frac{1}{2}, \ell\in\mathbb{R}$<br>$m-\ell\in\mathbb{N}$ |
| 3. | $m-1/2$ | $(\frac{1}{2}k^2)^2 j(j-m-\frac{1}{2})(2j-2m+2\ell)(2j+2\ell+1)$ | $\frac{k^2}{4}(2m^2+2m-\frac{1}{2})+\frac{(2\ell+1)^2}{4}+\frac{1}{2}(2-k^2)(2j-m+\frac{1}{2})$<br>$\times(2j-m+2\ell+\frac{3}{2})$ | $\dfrac{(\frac{k^2}{2})^j \prod_{i=1}^{m+3/2}(2\ell+2i+1)}{\prod_{i=1}^{m+3/2-j}(2\ell+2j+2i+1)}$ | $(\frac{i}{k'},\frac{-i}{k'})$ | $m\in\mathbb{N}-\frac{1}{2}, \ell\in\mathbb{R}$ |
| 4. | $m-1/2$ | ″ | ″ | ″ | $(\frac{-i}{k'},\frac{i}{k'})$ | $m\in\mathbb{N}-\frac{1}{2}, \ell\in\mathbb{R}$ |
| 5. | $\ell-1/2$ | $(\frac{1}{2}k^2)^2 j(j-\ell-\frac{1}{2})(2j-2\ell+2m)(2j+2m+1)$ | $\frac{k^2}{4}(2\ell^2+2\ell-\frac{1}{2})+\frac{(2m+1)^2}{4}+\frac{1}{2}(2-k^2)(2j-\ell+\frac{1}{2})$<br>$\times(2j-\ell+2m+\frac{3}{2})$ | $\dfrac{(\frac{-k^2}{2})^j \prod_{i=1}^{\ell+3/2}(2m+2i+1)}{\prod_{i=1}^{\ell+3/2-j}(2m+2j+2i+1)}$ | $(i,-i)$ | $\ell\in\mathbb{N}-\frac{1}{2}, m\in\mathbb{R}$ |
| 6. | $\ell-1/2$ | ″ | ″ | ″ | $(-i,i)$ | $\ell\in\mathbb{N}-\frac{1}{2}, m\in\mathbb{R}$ |
| 7. | | $(\frac{1}{2}k'^2)^2 j(j-n-1)(2j-2n-5)(2j+3)$ | $-\frac{k'^2}{2}(n+1)(n+4)-\frac{1}{2}(1+k^2)(n-2j)^2$ | $\dfrac{(\frac{k'^2}{2})^j \prod_{i=0}^{j+1}(2j+3-2i)}{3}$ | $(1,-1)$ | $n\in\mathbb{N}-1$ |
| 8. | | $\frac{1}{4}j(j-n-1)(2j-2n-5)(2j+3)$ | $-\frac{1}{2}(n+1)(n+4)-\frac{1}{2}(1-2k^2)(n-2j)^2$ | $\dfrac{(\frac{1}{2})^j \prod_{i=0}^{j+1}(2j+3-2i)}{3}$ | $(1,-1)$ | $n\in\mathbb{N}-1$ |

The coefficients $\rho_j$, $\lambda_j$, $\omega_j$ of the recurrence relation (4.17), (4.18) are given for different cases. The choice of roots $\xi_1$, $\xi_2$ and the restrictions on $m,\ell$ are also provided for each case.